\documentclass[acmsmall,screen, nonacm]{acmart}

\AtBeginDocument{%
  \providecommand\BibTeX{{%
    Bib\TeX}}}

\usepackage{fancyhdr}

\AtBeginDocument{%
    \addtolength{\footskip}{2.0\baselineskip}%
    \fancyfoot[L]{\textit{\textbf{Preprint --- Accepted to ISSTA 2025}}}%
}

\usepackage{siunitx}

\usepackage{amsmath,amsfonts}
\usepackage{algorithmic}
\usepackage{graphicx}
\usepackage{textcomp}
\usepackage{url}
\usepackage{listings}
\usepackage[shortcuts]{extdash}
\usepackage{xcolor}
\usepackage{fancyvrb}
\usepackage{svg}

\usepackage{mathtools, multirow,booktabs}

\definecolor{strawberry}{rgb}{.86, .64, .7}
\definecolor{lightgray}{rgb}{.9,.9,.9}
\definecolor{darkgray}{rgb}{.4,.4,.4}
\definecolor{purple}{rgb}{0.65, 0.12, 0.82}

\newcommand{\rev}[1]{\textcolor{black}{#1}}
\newcommand{\reva}[1]{\rev{#1}}

\newcommand{\all}[1]{\rev{#1}}

\lstdefinelanguage{JavaScript}{
  keywords={typeof, new, true, false, catch, function, return, null, catch, switch, var, if, in, while, do, else, case, break},
  keywordstyle=\color{blue}\bfseries,
  ndkeywords={class, export, boolean, throw, implements, import, this},
  ndkeywordstyle=\color{darkgray}\bfseries,
  identifierstyle=\color{black},
  sensitive=false,
  comment=[l]{//},
  morecomment=[s]{/*}{*/},
  commentstyle=\color{purple}\ttfamily,
  stringstyle=\color{red}\ttfamily,
  morestring=[b]',
  morestring=[b]",
  numbers=left,
  xleftmargin=0.5cm,
}

\def\BibTeX{{\rm B\kern-.05em{\sc i\kern-.025em b}\kern-.08em
    T\kern-.1667em\lower.7ex\hbox{E}\kern-.125emX}}

    
\begin{document}

% \title{Autonomous Web Tester Agents: Are We There Yet?}
\title{Are Autonomous Web Agents Good Testers?}

\author{Antoine Chevrot}
\email{antoine.chevrot@gmail.com}
\orcid{0000-0003-3677-5150}
\affiliation{%
  \institution{Smartesting}
  \city{Besançon}
  \country{France}
}

\author{Alexandre Vernotte}
\email{alexandre.vernotte@smartesting.com}
\orcid{0000-0002-2113-4900}
\affiliation{%
  \institution{Smartesting}
  \city{Besançon}
  \country{France}
}

\author{Jean-Rémy Falleri}
\email{jean-remy.falleri@u-bordeaux.fr}
\orcid{0000-0002-8284-7218}
\affiliation{%
  \institution{LaBRI}
  \city{Bordeaux}
  \country{France}
}

\author{Xavier Blanc}
\email{xavier.blanc@u-bordeaux.fr}
\affiliation{%
  \institution{LaBRI}
  \city{Bordeaux}
  \country{France}
}

\author{Bruno Legeard}
\email{bruno.legeard@smartesting.com}
\affiliation{%
  \institution{Smartesting}
  \city{Bensançon}
  \country{France}
}

\renewcommand{\shortauthors}{Chevrot et al.}

\begin{abstract}
Despite advances in automated testing, manual testing remains prevalent due to the high maintenance demands associated with test script fragility—scripts often break with minor changes in application structure. Recent developments in Large Language Models (LLMs) offer a potential alternative by powering Autonomous Web Agents (AWAs) that can autonomously interact with applications. These agents may serve as Autonomous Test Agents (ATAs), potentially reducing the need for maintenance-heavy automated scripts by utilising natural language instructions similar to those used by human testers.
\reva{This paper investigates the feasibility of adapting AWAs for natural language test case execution and how to evaluate them}. 

We contribute with (1) a benchmark of \all{three offline} web applications, and a suite of \all{$113$ manual} test cases, split between passing and \all{failing} cases, to evaluate and compare ATAs performance, (2) \rev{SeeAct-ATA and pinATA, two open-source ATA implementations capable of executing test steps, verifying assertions and giving verdicts}, and (3) comparative experiments using our benchmark that quantifies our ATAs effectiveness. Finally we also proceed to a qualitative evaluation to identify the limitations of PinATA, our best performing implementation.

\rev{Our findings reveal that our simple implementation, SeeAct-ATA, does not perform well compared to our more advanced PinATA implementation when executing test cases ($50\%$ performance improvement). However, while PinATA obtains around $60\%$ of correct verdict and up to a promising 94\% specificity, we identify several limitations that need to be addressed to develop more resilient and reliable ATAs, paving the way for robust, low maintenance test automation.}
\end{abstract}


\maketitle

\section{Introduction}
\label{intro}

In the ever-evolving landscape of web development, ensuring the quality and reliability of web applications is a critical endeavour. 
Traditionally, application testing has relied heavily on manual processes, particularly for end-to-end test cases where the system under test is a web application. 
While these manual approaches are effective, they have significant drawbacks regarding time, cost, and scalability. 
As web applications grow in complexity and the demand for faster development cycles intensifies, the limitations of manual testing become more apparent. 
Manual tests are not only time-consuming to \rev{execute}, but also prone to human error and inconsistency. 
Furthermore, the cost of maintaining a large team of manual testers can be prohibitive for many organisations, especially as applications require more frequent updates and releases, which is the case of web applications.

To overcome these limitations, test automation was introduced, with the main objective of providing executable test code. 
In the context of web applications, several frameworks and tools have been proposed to facilitate the production of this code, from automatic browser manipulation to automated test code generated by recorded user behaviour. 
Today, however, test automation fails to live up to its promise, mainly because automated test \rev{scripts} are too often fragile and require significant maintenance 
effort. This conclusion for web applications is observed in both Record/Replay tests~\cite{hammoudi_why_2016}, and end-to-end tests~\cite{memon_chapter_2019, leotta_challenges_2023}. 
We are even witnessing the emergence of methods for estimating the cost of maintenance and thus measuring the benefits of deploying full test automation~\cite{di_meglio_towards_2024}. 

Artificial intelligence (AI), particularly the generative capabilities of large language models (LLMs), has demonstrated potential for generating automated test \rev{scripts}~\cite{yang_evaluation_2024,yu_llm_2023,wang_software_2024}. The primary goal is to reduce the time required to create automated tests and to minimise errors. Additionally, LLMs provide valuable support in generating CSS selectors, a common source of test fragility. However, while these advancements enhance the efficiency of test code production, the generated code remains specific to a particular version of the target application, leaving test fragility unresolved.

Recently, Autonomous Web Agents (AWA), which leverage LLMs and integrate the capability to interact with web applications~\cite{wang_survey_2024}, have demonstrated the potential to perform tasks autonomously~\cite{zhang_webpilot_2024, iong_openwebagent_2024, he_webvoyager_2024, chen_webvln_2023}, as described in Section~\ref{sec:background}.
This development raises an intriguing question: could an AWA execute test \rev{cases}, effectively taking over the role of manual testers in web application testing? 

Manual test instructions closely resemble the input formats that AWAs use, and their broad prior knowledge may allow these agents to be more naturally resilient to minor changes in web application design compared to traditional test code.
\rev{This is therefore not a surprise that we are now witnessing the emergence of autonomous testing agents in commercial products such as Flowtest.ai or Octomind.dev.
However, while a few products exist, little is known about how these autonomous testing agents work and what results they are capable of, raising the question of their concrete interest.}

\rev{In this article, we seek to answer these questions. Since there is no prior literature on how to build an autonomous testing agent, we leverage the literature on AWAs to create and implement two alternative designs for a test agent, presented in  Section~\ref{sec:ata}: a baseline design and a more advanced design where we select and adapt to the testing context state-of-the art concepts from web agents.}

\rev{To evaluate these designs on their ability \rev{to execute test cases} on web applications, we create a fully reproductible benchmark, presented in Section~\ref{sec:benchmark}, containing web applications and manual test cases written by four professional testers (i.e., written in natural language) and define metrics to evaluate the ability of an LLM agent to execute these tests automatically. This benchmark is built with extensibility in mind and we hope that it will help driving the research in the area of autonomous test agents.}

\rev{Finally, we use our benchmark to evaluate our two alternative designs and analyse their performances though a quantitative study, described in Section~\ref{sec:experiments}. We also proceed to a qualitative assessment of the behaviour of agents to identify the pain points of our agents, yielding future research directions for the community.}

Our results show that it is possible to transform autonomous agents from the web automation domain so that they can automatically execute tests without any prior knowledge of the application to be tested: \rev{more than the half} of the tests of our benchmark are successfully run by the agent as they were run by a manual tester.
However, a range of issues still prevent web agents from reliably performing end-to-end tests, notably, their inclination toward misinterpreting actions or assertions.

\section{Test and Automation for Web Applications}
\label{sec:background}

This section begins by introducing an example of an end-to-end (E2E) test \rev{case}, which will serve as a continuous reference throughout the article. The example is used to illustrate key concepts and methodologies discussed in subsequent sections. The section proceeds by demonstrating how the E2E test \rev{case} can be automated by generating executable code, leveraging a framework like Playwright~\footnote{\url{https://playwright.dev/}} to facilitate browser manipulation and interaction. Finally, the section concludes with a discussion on the use of autonomous agents, specifically designed for autonomous task execution on the web, and explores how these agents can be utilised to automate and execute test cases effectively.

\subsection{Manual E2E Testing}

\begin{figure*}
    \centering
    \includegraphics[width=\textwidth]{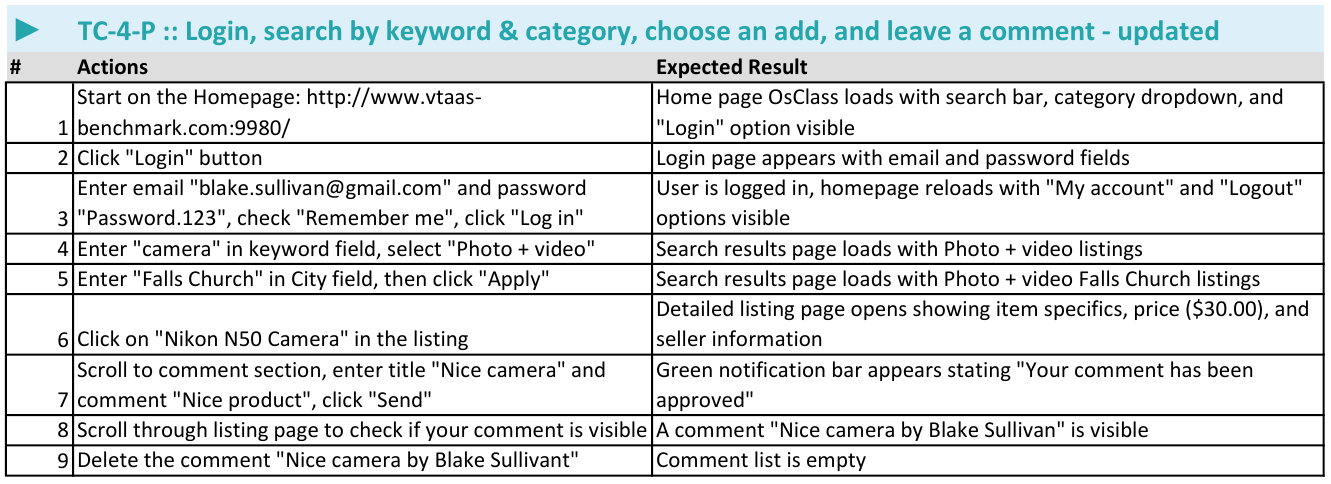}
        \captionsetup{
    width=.95\linewidth,      % Width same as table
    justification=centering,
    singlelinecheck=false,  % Ensures centering works for multi-line captions
    font=small,            % Match table font size
    labelfont=bf,          % Bold "Table X"
    textfont=normal        % Normal text for caption
}
    \caption{An E2E test \rev{case} for the Classified Application}
    \label{fig:tc}
    \Description{An E2E test \rev{case} for the Classified Application}
\end{figure*}

An E2E test \rev{case} simulates a complete user scenario to verify that the workflow functions as expected within a web application. It describes the scenario sequentially, specifying the test actions required at each step, as well as any \rev{assertions} to validate the outcome by observing the application's state. For simplicity, an E2E test \rev{case} can be represented as a two-column table: the left column outlines the test actions, and the right column lists the \rev{test assertions}. Each row corresponds to a step in the scenario, and it is possible for some steps to omit \rev{test assertions}. Figure~\ref{fig:tc} illustrates an example of an E2E test \rev{case} designed for a marketplace application\footnote{\url{https://osclass-classifieds.com/}}. 
The objective of this nine-step test is to ensure that a user can successfully log into the application, search for a product using a keyword, select the product, and add a comment.

E2E test \rev{cases} are frequently performed manually by human testers. At each step, the tester must carry out the specified test action and verify that the observed result matches the expected result. If the tester is unable to execute an action or confirm the expected result for any step, that step is marked as a failure, and consequently, the entire test is considered to have failed. Conversely, if the tester successfully completes all steps, performing all test actions and confirming that the results match the expected results, the test is marked as passed.

\subsection{E2E Testing Automation}

The manual execution of E2E test \rev{cases} can be both time-consuming and prone to errors. To address these challenges, the automation of E2E tests has been explored as a solution~\cite{dobles_comparing_2019}. Automating an E2E test involves generating executable code to carry out all the steps defined in the test scenario. This includes automating both the test actions with the web application and the verification of expected results. To facilitate this process, several frameworks specifically designed for web application automation have been developed~\cite{garcia_exploring_2024,anupam_automating_2000}.

The code in Listing~\ref{fig:test-code} demonstrates the automation of the first three steps of the E2E test \rev{case} illustrated in Figure~\ref{fig:tc}. 
In this example, the Playwright web automation framework is used to interact with the web application. For instance, line 16 shows how a click event on the login button is automated. Additionally, the code demonstrates how \rev{test assertions} are implemented. 
For example, lines 11, 12, and 13 validate the presence of three critical web elements—the search bar, category drop-down list, and login options, which confirm that the homepage has successfully loaded.

\begin{lstlisting}[language=Javascript,caption=Excerpt of the source code automating the test case presented in Figure~\ref{fig:tc},label={fig:test-code}]
// Step 1: Start on the Homepage
await page.goto('http://www.vtaas-benchmark.com:9980/');

// Verify homepage loaded correctly
await page.waitForSelector('input[name="sPattern"]'); // Search bar
await page.waitForSelector('select[name="sCategory"]'); // Category dropdown
await page.waitForSelector('a:has-text("Login")'); // Login option

// Step 2: Click "Login" button
await page.click('a:has-text("Login")');

// Step 3: Enter email, password, check "Remember me", click "Log in"
await page.fill('input[name="email"]', 'blake.sullivan@gmail.com');
await page.fill('input[name="password"]', 'Password.123');
await page.check('input[name="remember"]');
await page.click('button:has-text("Log in")');

// Verify user is logged in
await page.waitForSelector('a:has-text("My account")');;
\end{lstlisting}

It is important to note that the source code of an automated test is inherently tied to a specific version of the web application being tested. This dependency arises from assumptions the test makes about the Document Object Model (DOM) structure of the application, which are used to uniquely identify the web elements with which it interacts. For instance, on line 12, the test assumes that the search bar can be located using the CSS selector \texttt{input[name="sPattern"]}. Similarly, on line 16, it relies on the assumption that the login button can be identified by the CSS selector \texttt{a
("Login")}.

These assumptions render automated tests fragile, as they are often invalidated by changes in the web application's structure over time~\cite{hammoudi_why_2016}. For example, the CSS selectors for the search bar and login button may be altered in future versions of the application. As a result, maintaining the automated test code to accommodate these changes can be both time-consuming and costly~\cite{leotta_challenges_2023}, introducing significant overhead in terms of maintenance and additional resource requirements.

\subsection{Autonomous Web Agents}
Recent advancements in the domain of autonomous agents present promising opportunities for addressing long-standing challenges in test automation, particularly regarding the fragility of test code. An autonomous agent is an entity designed to perform tasks independently, capable of interacting with its environment through actions, and utilising the capabilities of LLMs to interpret its surroundings without prior knowledge~\cite{wang_survey_2024}. The agent determines the necessary actions to achieve its assigned task based on this understanding. In the context of web applications, this entails executing tasks such as ordering an item from an online store, booking a trip, or searching for information on a website.

\begin{figure*}
    \centering
    \includegraphics[width=0.7\textwidth]{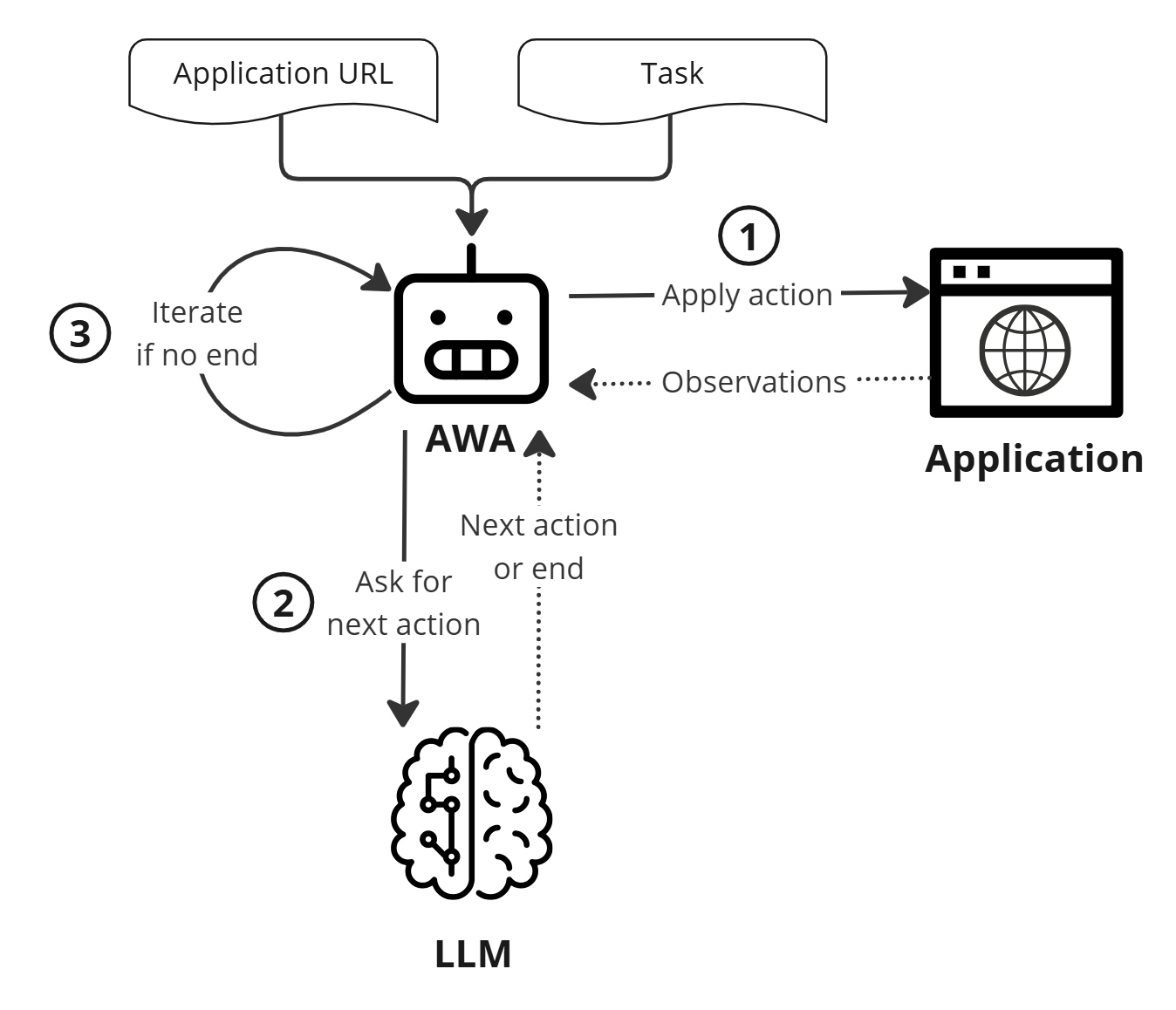}
    \captionsetup{
    width=.95\linewidth,      % Width same as table
    justification=centering,
    singlelinecheck=false,  % Ensures centering works for multi-line captions
    font=small,            % Match table font size
    labelfont=bf,          % Bold "Table X"
    textfont=normal        % Normal text for caption
}
    \caption{Autonomous Web Agent main \reva{process}}
    \Description{Autonomous Web Agent main process}
    \label{fig:awa}
\end{figure*}

\rev{Existing approaches that implement AWAs typically take a task description in natural language and the base URL of the web application as input. These approaches generally follow a structured iterative process as presented in the \ref{fig:awa}. In each iteration, (1) the agent performs the most appropriate action on the web application (in the first iteration, this action involves opening the application at the base URL) and then observes the environment by capturing some elements such as screenshots and/or the DOM source code; (2) the agent analyses these elements, interpret the task requirements, and infer the next appropriate action or determine whether the task has been completed. To do this, the agent constructs one or several prompts that include relevant questions for the LLM, as well as all necessary data, such as observations, task details, and previously performed actions; (3) based on the LLM's response, the agent decides whether to initiate a new iteration or terminate the process.}

AWA approaches offer significant potential for applications in test automation for web applications. In this context, the task assigned to the agent can be interpreted as executing a test scenario. For example, an agent could be tasked with executing the test outlined in Figure \ref{fig:tc}. The agent would autonomously analyse the test, observe the application to identify the necessary actions, and perform the required \rev{assertions}. This approach enhances the robustness of the test automation process by reducing the vulnerability of changes in the web application’s structure.

\section{From Autonomous Web Agent to Autonomous Test Agent}
\label{sec:ata}

\rev{This section starts by highlighting the key differences between AWA and ATA. 
It then presents the 4 main modules of a classical AWA architecture and details the main modifications that should be applied to them for transforming an AWA into an ATA. 
Finally, it presents two ATAs we developed.
The first one, SeeAct-ATA, is a basic ATA that relies entirely on a LLM and is based on the SeeAct AWA~\cite{zheng_gpt-4vision_2024}. 
The second, PinATA, is a more advanced ATA, incorporating state-of-the-art techniques for perception, reasoning, evaluation, and grounding~\cite{qin_ui-tars_2025}.}

\subsection{Main Differences Between AWAs and ATAs}

While the execution of an E2E test \rev{case} can be considered a task in its own right, several unique features differentiate it from conventional tasks, requiring modifications to an AWA to transform it into an Autonomous Test Agent (ATA).

The first critical distinction between an E2E test \rev{case} and a typical task is the strict sequence of actions that must be followed. An E2E test \rev{case} outlines a predefined scenario consisting of a series of steps that must be executed in the correct order, like in Figure~\ref{fig:tc}. Unlike an AWA, which focuses on achieving the overall task objective, regardless of the actions taken or their order, an ATA is constrained by the test scenario and must strictly adhere to the specified sequence of steps.

The second key feature of E2E testing is the requirement for multiple intermediate \rev{assertions} throughout the process. If any of these \rev{assertions} fail, the entire test is deemed unsuccessful. This behaviour differs from that of standard AWAs, which aim to complete the task without performing intermediate validations.
Despite these two key differences, ATAs and AWAs share many fundamental similarities. Both agents must be capable of executing actions, observing their environment, and using this information to guide their decisions.

\subsection{AWA Architectures}

\rev{Wang et al.~\cite{wang_survey_2024} define the architecture of an AWA as consisting of four fundamental modules: profile, memory, planning, and action. The profile module specifies how the agent interacts with the LLM, including its role and self-description, all of which influence the LLM’s responses~\cite{sun_building_2024}.}
\rev{The memory module manages the information necessary for the task execution, encompassing both data collected from observing the application and external knowledge retrieved from the internet.}
\rev{The planning module serves as the agent’s reasoning component, enabling decision-making for selecting subsequent actions. Finally, the action module defines the set of tools available to the agent to interact with the application and its broader execution environment.}

\rev{Adapting an AWA into an ATA requires modifications to all four modules. Specifically, the profile module must explicitly define the agent’s role as a tester, with the primary objective of executing a predefined test scenario.
The memory module should retain a record of all performed test steps, as future steps may rely on this information.
For example, certain tests require verifying that a value obtained at a specific step matches one retrieved earlier; the memory module should facilitate such verifications.
The planning module must strictly enforce adherence to the test scenario, ensuring that all actions are executed in the specified order.}
\rev{The action module should not only support the test actions with the application but also the performing of assertions.
The two following subsections present how we realised these modifications to build two ATAs.}

\subsection{SeeAct-ATA}

\rev{We first choose to develop a very basic ATA that serves as a baseline.
We opted to develop it by adapting the SeeAct AWA~\cite{zheng_gpt-4vision_2024} since it uses a very elementary architecture where the 4 main modules are encoded into a prompt, relying mainly on the LLM for most of the work, making it easy to adapt as an ATA.}
\rev{This prompt contains multiple questions to guide the LLM in proposing the next action required to complete the given task.} A simplified version of this prompt is presented in Listing \ref{lst:prompt-awa}. The prompt begins by defining a human user profile. Next, relevant data is integrated into the prompt, including the task to be completed, the previous actions undertaken, and the current screenshot of the application. 
Following this, several instructions are provided, utilising the Chain of Thought~\cite{wei_chain--thought_2023} approach -- prompting process by delineating complex tasks into a sequence of logical steps towards a final resolution --, which has been shown to enhance the LLM’s reasoning and output quality. While the full set of instructions is not detailed here, the prompt’s goal is to have the LLM describe the web page, summarise the previous actions and the screenshot, and then propose the next action. The LLM is also instructed to align its proposed action with the interactive elements identified in the DOM, as provided in the observations. Finally, the LLM returns the result in a specific format. The action 'TERMINATE' allows the LLM to indicate that the task has been completed. \rev{If not completed, the LLM returns the next action to be performed. The SeeAct AWA then relies on Playwright to perform this action.} 

\begin{lstlisting}[caption=Prompt used by seeAct to ask the LLM for the next step,label={lst:prompt-awa},captionpos=b]
Imagine that you are imitating humans doing web navigation. You need to decide on the first following action to take. 

You are asked to complete the following task: {{ TASK }} 

Previous Actions: {{ PREVIOUS_ACTIONS }}

The screenshot below shows the webpage you see. {{ SCREENSHOT }}

Follow the following guidance:

[CURRENT WEBPAGE IDENTIFICATION]: Firstly, think about what the current webpage is.

[PREVIOUS ACTION ANALYSIS]: Secondly, combined with the screenshot, analyse each step of the previous action history and their intention one by one.

[SCREENSHOT DETAILS ANALYSIS]: Closely examine the screenshot to check the status of every part of the webpage to understand what you can operate with and what has been set or completed. 

[NEXT ACTION BASED ON WEBPAGE AND ANALYSIS]: Then, based on your analysis, decide on the following action. And clearly outline which element will operate with as the first next target element, its detailed location, and the corresponding operation.

[MULTICHOICE QUESTION]: Below is a multi-choice question, where the choices are elements in the webpage.  Determine whether one matches your target element.
If none of these elements match your target element, please select N. None of the other options match the correct element.
A. <a>Jump to sidebar</a>
B. <a>Jump to main content</a>
C. <a>Go to HomePage</a>
[...]
N. None of the other options match the correct element

[FINAL ANSWER]: Finally, conclude your answer using the format below.
    ELEMENT: The uppercase letter of your choice. (No need for PRESS ENTER)
    ACTION: Choose an action from {CLICK, SELECT, TYPE, PRESS ENTER, TERMINATE, NONE}.
    VALUE: Provide additional input based on ACTION.
\end{lstlisting}

To transform SeeAct into an ATA, we modified its prompt to instruct the LLM to propose the next action according to the specifications of the test scenario and to perform verification assertions.
Listing~\ref{lst:prompt-ata} highlights the key differences between the original SeeAct prompt and the revised version we developed to adapt it into an ATA.
First, we defined a new profile representing a manual tester to better align with a test execution mindset.
Next, we set the E2E test \rev{case} as the task to be executed. Additionally, we introduced two new instructions:
\begin{itemize}
    \item \emph{Test Case Progress:} This instruction directs the LLM to identify which steps of the test have already been executed and determine the current step. This helps focus the LLM on the specific part of the test that needs to be executed.
    \item \emph{Test Step Assertion Control:} This instruction asks the LLM to perform the assertions specified in the current test step.
\end{itemize}
 
The rest of the prompt remains unchanged from the original SeeAct configuration that can be found on their Github repository~\footnote{\url{https://github.com/OSU-NLP-Group/SeeAct}}.

\begin{lstlisting}[caption={Our ATA prompt with main differences with the AWA prompt highlighted in blue},label={lst:prompt-ata},captionpos=b]
<@\textcolor{blue}{Imagine that you are imitating a manual tester performing a test case. 
You receive a test case and you should follow it and determine its verdict. }@>

You are asked to complete the following task: <@\textcolor{blue}{\{\{ TEST \}\} }@>

Previous Actions: {{ PREVIOUS ACTIONS }}

The screenshot below shows the webpage you see.

[CURRENT WEBPAGE IDENTIFICATION]: ...

[PREVIOUS ACTION ANALYSIS]: ...

<@\textcolor{blue}{[TEST CASE PROGRESS]: Based on the past actions, determine the state of the test case. 
Go through each step, one by one, and give it a status: {DONE, CURRENT, TODO}. 
Finish by confirming the current test step.}@>

[Screenshot Details Analysis]: ...

<@\textcolor{blue}{[TEST STEP ASSERTION CONTROL]: based on past actions, if any, control the test step assertion. Make a numbered list of all the atomic assertions and, for each one, based on what you see in the screenshot, assign a status {VERIFIED, NOT VERIFIED}. If all assertions are verified, write the sentence: All assertions have been verified}@>

[NEXT ACTION BASED ON WEBPAGE AND ANALYSIS]: ...

[MULTICHOICE QUESTION]: ...

[FINAL ANSWER]: ...
\end{lstlisting}

\rev{Our transformation of SeeAct into an ATA, named SeeAct-ATA, is intentionally basic and minimal to evaluate the feasibility of using autonomous agents in a test automation context.}
This streamlined adaptation allows us to explore how effectively autonomous agents can handle test cases while maintaining the core functionality of the original AWA architecture.

\subsection{PinATA}

\rev{Our second ATA, called PinATA for \textbf{P}lanned \textbf{IN}centive ATA  is more advanced than the first one.
We picked up from the literature approaches that we believe are well suited to comply with the specific objectives of an ATA.
Unlike our first ATA, which relies on a single prompt, PinATA is composed of three specialised components: the orchestrator, the actor, and the assertor.}

\rev{The orchestrator oversees the entire process, delegating action execution to the actor and verification tasks to the assertor.}
\rev{According to the four modules architecture, the orchestrator corresponds to the planning module, while the actor and assertor together form the action module. They all share a global memory module and further have a similar profile module.}

\rev{The orchestrator is leading the process employing a \emph{Planning with Feedback} strategy~\cite{wang_survey_2024}.
It maintains a model of the test scenario and monitors its execution progress. At each step, it instructs the actor to perform the corresponding action. Based on the actor’s feedback, the orchestrator determines whether the action was successfully executed. If not, it requests a retry until either the action is completed or deemed unfeasible, in which case the test is marked as failed.}
\rev{Once an action is executed, the orchestrator delegates the verification task to the assertor. Similarly, it evaluates the assertor’s feedback to determine whether the assertion was successfully performed. If necessary, it requests a retry. Depending on the outcome of the verification, the orchestrator either proceeds to the next step or concludes that the test has failed.}

\rev{The actor, tasked with executing actions, identifies the relevant elements of the target application for the test action—an operation known as \emph{grounding}, which remains a key challenge in AWA research~\cite{chandu_grounding_2021, gu_dont_2023}. Our implementation leverages the \emph{Set-Of-Marks} technique~\cite{yang_set--mark_2023} and uses the LLM’s capability to determine X and Y coordinates from screenshots. The actor component uses the Playwright framework for interacting with the browser.}

\rev{The assertor is responsible for performing verifications, employing the \emph{Agent as a Judge} approach to harness the LLM’s evaluative capabilities~\cite{zhuge_agent-as--judge_2024}. Our implementation asks the LLM to perform an evaluation just by analysing the screenshot of the application.}

\rev{All three components (orchestrator, actor and assertor) share a global memory system that follows a \emph{long-term memory} approach for the memory module~\cite{zhang_survey_2024}, storing all processed information. Notably, the system does not retrieve external data from the internet; it relies solely on observations of the target application.}

\rev{Finally, all three agents share a similar profile module, presenting themselves as part of a multi-agent system while clearly defining their specific roles.}

\rev{PinATA incorporates what the authors' consider to be a representative selection of state-of-the-art approaches from existing AWAs. However, we recognise that the design space for ATAs is vast, necessitating systematic evaluation methodologies. To address this need, we introduce in the following section a comprehensive benchmark designed to enable objective comparison of different ATA implementations.}

\section{A Benchmark for assessing ATA effectiveness}
\label{sec:benchmark}
This section presents the fully reproducible benchmark developed to objectively evaluate ATA effectiveness in executing E2E test \rev{cases} originally designed for manual execution. It begins by outlining the motivations for defining a new benchmark, followed by presentation of the three web applications serving as the system under test and the test cases. It ends with a description of the metrics used to measure effectiveness.

% Despite this distinction, we drew valuable insights from existing frameworks. We observed that web applications in current benchmarks are either public (production) applications or packaged (deployable) applications. We chose the latter approach using WebArena's applications to prioritize reproducibility, although this slightly sacrifices realism. Following best practices from AWA benchmarks, we incorporated test cases written by professional testers to ensure realism. Finally, we developed new metrics specifically tailored to measure ATA effectiveness, as existing AWA metrics focus on task completion rather than test execution.
% Our benchmark has been developed following established guidelines~\cite{v_kistowski_how_2015}, ensuring its suitability and replicability for future studies. It includes realistic and fully deployable web applications, including code and data to enable reproducibility. It includes realistic passing and failing end-to-end test cases associated with their ground-truth verdict. It defines a comprehensive set of metrics that allows the assessment and comparison of ATA effectiveness.

\subsection{Why Yet Another Benchmark?}
\rev{Several benchmarks have been developed to evaluate AWA effectiveness. Famous examples include Mind2Web, which defines 2,350 tasks across 137 websites to assess how well AWAs can accomplish them~\cite{deng_mind2web_2023}; WebArena, which provides a standalone, self-hostable web environment with hundreds of tasks to measure AWA performance~\cite{zhou_webarena_2024}; the TheAgentCompany, which provides more complex tasks~\cite{xu_theagentcompany_2024}; and AgentQuest, which offers a similar benchmark but incorporates more flexible evaluation metrics~\cite{gioacchini_agentquest_2024}. Qin et al.~\cite{qin_ui-tars_2025} provide a comprehensive analysis of existing benchmarks designed for AWA evaluation.}

\rev{All of these benchmarks define a set of tasks that AWAs must complete, making them unsuitable for evaluating ATAs. As discussed in Section \ref{sec:background}, ATAs take as input a test case written in natural language, which consist of sequential steps and are not directly comparable to standalone tasks. This fundamental difference motivated the development of a new benchmark specifically designed for ATA evaluation.}

\rev{We observed that web applications used in existing benchmarks are either public applications, which are in production and thus subject to uncontrollable modifications, or packaged applications, which can be deployed as-is and fixed in time.} %Public applications provide a more realistic environment but are subject to frequent changes, potentially misaligning with the benchmark’s requirements and compromising replicability. In contrast, packaged applications, while slightly less realistic, enable the creation of a fully replicable benchmark. 
\rev{Given the importance of reproducibility, we opted for the latter approach and leveraged three applications provided by the WebArena benchmark.}

\rev{AWA benchmarks also highlight the importance of using real-world data, particularly for the tasks they evaluate. Following this principle, we incorporated test \rev{cases} written by professional testers into our benchmark to ensure its realism and applicability.}

\rev{Finally, we developed new metrics specifically tailored to measure ATA effectiveness, as existing AWA metrics focus on task completion rather than test execution.}

\rev{As a summary, our benchmark has been developed following established guidelines~\cite{v_kistowski_how_2015}, ensuring its suitability and replicability for future studies. It includes realistic and fully deployable web applications, including code and data to enable reproducibility. It includes realistic passing and failing end-to-end test cases associated with a verdict given by expert testers. It defines a comprehensive set of metrics that allows the assessment and comparison of ATA effectiveness.}

\subsection{Web Applications}

% For our ATA benchmark, we selected three web applications previously utilised in the Visual Web Arena~\cite{zhou_webarena_2024}, known for supporting web automation research. Each provides a unique environment with diverse test scenarios:

Our benchmark contains three web applications previously utilised in the Visual Web Arena benchmark~\cite{zhou_webarena_2024}. Each provides a unique environment for diverse test cases:

\begin{itemize}
    \item \textbf{Classifieds}: Inspired by real-world marketplaces like Craigslist and Facebook Marketplace, this site includes 65,955 listings and offers a distinct environment focused on marketplace-style user test actions.
    \item \textbf{Postmill}: A social networking site inspired by Reddit, allowing users to create comments on various topics organised into subsections.
    \item \textbf{OneStopShop}: A shopping site featuring extensive search filters for a wide array of products, with capabilities for account creation, shopping cart management, and order placement.
\end{itemize}

\all{This choice, as outlined in the previous subsection, is driven by our primary goal of creating a fully reproducible benchmark. Each web application is provided as a Docker image, allowing for easy deployment. Even though the web applications of our benchmark target popular domains (marketplaces, social
networking, and e-commerce), incorporating a larger set of web applications would further strengthen
the benchmark. However, we feel that this first core set of applications constitutes a good first milestone to evaluate ATA. We hope that open-sourcing the benchmark and the models will allow the growth of this core through community-driven efforts.}% adding a new web application is work intensive: it requires identifying a website with its associated data and creating dedicated tests for it as described in the next section. For this reason we limited ourselves to three applications, but made our benchmark fully open-source with the hope to enable future contributions.}

\subsection{Test Cases}

\rev{We engaged four professional testers to design test cases for the three web applications in our benchmark. Each tester has over five years of experience in web application testing and holds an ISTQB~\footnote{\url{https://www.istqb.org/}} Foundation Level certification. 
We instructed them to design passing test cases and document them using a two-column matrix, as illustrated in Figure~\ref{fig:tc}. Each test case comprises multiple steps, detailing specific actions along with optional assertions. Each application was assigned a dedicated tester, except for the \textit{Classified} application, which was tested by two professionals. Over three weeks, the testers designed a total of 62 test cases.}

\rev{After receiving the test cases, we applied several post-processing steps to ensure consistency and reliability. First, we verified that all test cases were self-contained, including any required data (e.g., login credentials) to facilitate successful execution. We assumed that each test case would be executed on a freshly deployed web application to prevent interference from previous test runs. Additionally, we ensured that all test cases were written in English to enhance the benchmark’s usability. Finally, one of the authors manually re-executed each test case to validate its correctness.}

\rev{Considering the time required for test case design by professional testers—along with additional efforts such as translation and manual execution—each successfully passing test case represents an average investment of approximately one hour.}

\begin{figure*}
    \centering
    \includegraphics[width=0.9\textwidth]{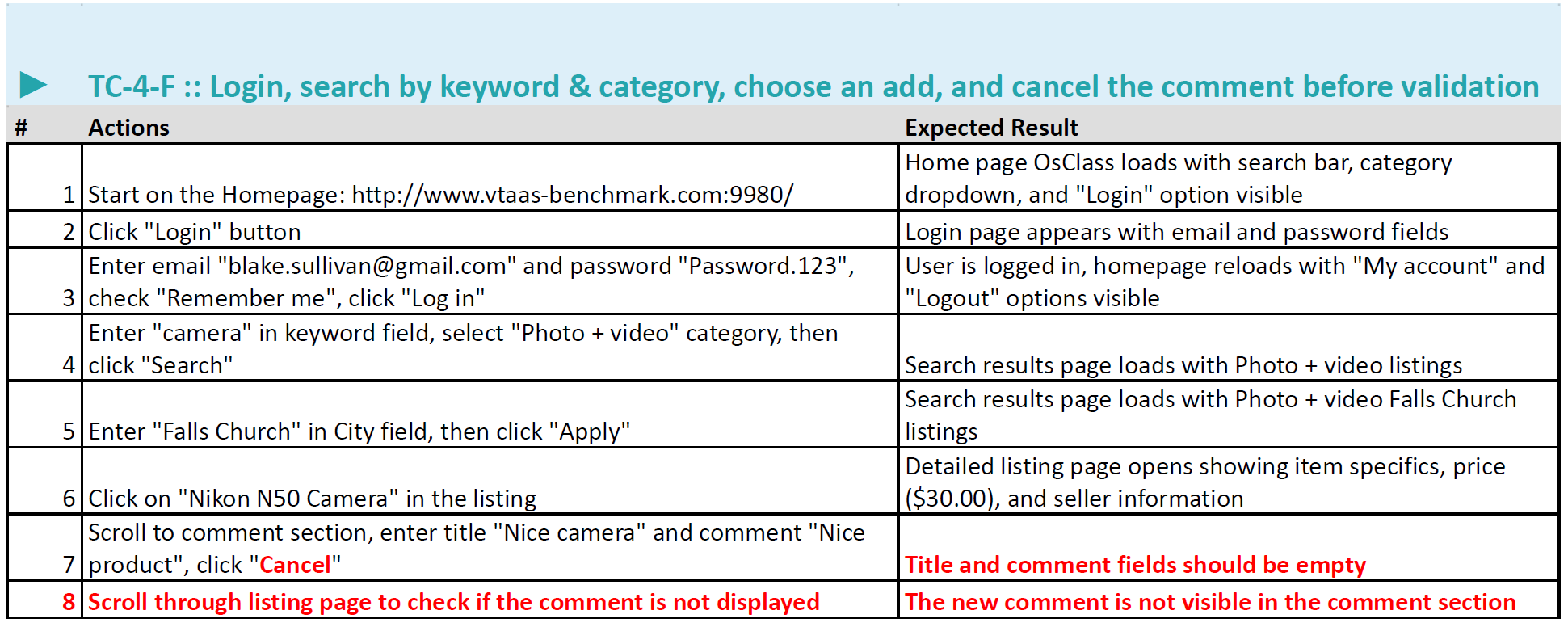}
    \Description{A failing test case obtained by modifying the test case presented in fig~\ref{fig:tc}. Changes appear in bold red.}
        \captionsetup{
    width=.95\linewidth,      % Width same as table
    justification=centering,
    singlelinecheck=false,  % Ensures centering works for multi-line captions
    font=small,            % Match table font size
    labelfont=bf,          % Bold "Table X"
    textfont=normal        % Normal text for caption
}
    \caption{\all{A failing test case obtained by modifying the test case presented in fig~\ref{fig:tc}. Changes appear in bold red.}}
    \label{fig:failing}
\end{figure*}

\rev{To integrate non-passing test cases into our benchmark and ensure a balanced dataset, two authors collaborated to derive failing test cases from existing passing ones. One of these authors is an experienced professional tester who verified the correctness and industry compliance of the generated tests.
Failing test cases were constructed by modifying passing ones to check for non-implemented features within the application. As shown in Figure~\ref{fig:failing}, we adapted the passing test case from Figure~\ref{fig:tc} to create a corresponding failing version. In this example, the cancel button is absent, making it impossible to click. A correctly functioning ATA should halt execution at this point rather than proceeding further. Each failing test case includes an explicitly annotated expected failure step, enabling precise metric calculations during execution.
Out of the 62 passing test cases, 51 corresponding failing test cases were developed. The discrepancy arises because some passing tests were highly similar, leading us to generate a single failing test case for such groups. For instance, the passing tests P19, P20, and P21 in OneStopShop take place on the same web page and differ only slightly in their final actions. Instead of creating separate failing test cases for each, we produced a single representative failing test case.}

%\all{These failing tests simulates tests that were not updated along the application or, on the contrary, specification-compliant tests that are not reflected in the application. Either way, they are crucial for an ATA to correctly identify them as non-passing tests.}

\all{Our benchmark ultimately comprises 62 passing test cases, crafted by professional testers, and 51 failing test cases manually derived from the passing test cases. These test cases are distributed across the three selected websites, as detailed in Table~\ref{tab:initial_data}.}

\begin{table}[htbp]
\centering
\begin{tabular}{|l|c|c|c|}
\hline
\textbf{Application} & \textbf{\all{Application Domain}} & \textbf{Passing Tests} & \textbf{Failing Tests} \\
\hline
Classified  & \all{Shopping} & 15 & 15 \\
\hline
Postmill  & \all{Social Network} & \all{18} & \all{16} \\
\hline
OneStopShop  & \all{Flea Market} & \all{29} & 20 \\

\hline
\end{tabular}
\vspace{10pt}
\captionsetup{
    width=.95\linewidth,      % Width same as table
    justification=centering,
    singlelinecheck=false,  % Ensures centering works for multi-line captions
    font=small,            % Match table font size
    labelfont=bf,          % Bold "Table X"
    textfont=normal        % Normal text for caption
}
\caption{The distribution of the \all{$113$} test cases split among the \all{$3$} chosen websites}
\label{tab:initial_data}
\vspace{-20pt}
\end{table}
\subsection{Metrics}
\label{sec:metrics}

In this section, we present a set of metrics for evaluating the performance of ATAs against human testers.
Our metrics focus on two key aspects: verdict alignment and step alignment.

\paragraph*{Verdict Alignment}

Our metrics are defined based on evaluating an ATA's judgment as a binary classification problem. This approach models test case evaluation by defining two classes:
\begin{itemize}
\item \textit{Negative class:} a test case passes according to manual evaluation,
\item \textit{Positive class:} a test case fails according to manual evaluation.
\end{itemize}

To assess the alignment between human and agent verdicts, we define the following outcomes:

\begin{itemize}
\item \textit{True Positive (TP):} Both human and agent identify the test case as failing. This represents the correct identification of a failing test case.
\item \textit{True Negative (TN):} Both human and agent identify the test case as passing. This represents the correct identification of a passing test case.
\item \textit{False Positive (FP):} The agent identifies a test case as failing, but the human evaluator deems it passing. This is a Type I error, potentially leading to unnecessary debugging efforts.
\item \textit{False Negative (FN):} The agent identifies a test case as passing, but the human evaluator finds it failing. This is a Type II error, potentially allowing bugs to go undetected.
\end{itemize}

We then use a set of traditional metrics to evaluate how the results from an agent align with the manual results:

\noindent\begin{minipage}{.32\linewidth}
\begin{equation*}
\text{Accuracy} = \frac{\text{TP} + \text{TN}}{\text{TP} + \text{TN} + \text{FP} + \text{FN}}
\end{equation*}
\end{minipage}%
\begin{minipage}{.32\linewidth}
\begin{equation*}
\text{Specificity} = \frac{\text{TN}}{\text{TN} + \text{FP}}
\end{equation*}
\end{minipage}
\begin{minipage}{.32\linewidth}
\begin{equation*}
\text{Sensitivity} = \frac{\text{TP}}{\text{TP} + \text{FN}}
\end{equation*}
\end{minipage}\\

Accuracy provides an overall measure of how often the agent's verdict aligns with the human judgment.
To provide a more nuanced evaluation, we also consider 
specificity that measures the agent's ability to correctly identify passing test cases. A high specificity indicates that the agent is effective at avoiding false alarms.
Finally, sensitivity, also known as recall, assesses the agent's performance in detecting failing test cases. A high sensitivity suggests that the agent is effective at identifying discrepancies between test case and application when they are present.

\paragraph*{Step Alignment}

Note that an agent can identify a test as failing similarly to a human, but not for the same underlying reason (i.e. not failing on the same assertion).
This would be a type III error~\cite{schwartz_right_1999} where the verdict is correct but for a wrong reason.
To account for this possible behaviour, we define a second set of metrics based on the recorded step index where the failure is observed.
We decompose the true positives into three categories :

\begin{itemize}
    \item \textbf{AFB}: \textbf{A}gent \textbf{F}ails \textbf{B}efore human represents the number of cases where the agent fails at an earlier step than the human.
    \item \textbf{AFA}: \textbf{A}gent \textbf{F}ails \textbf{A}fter human represents the number of cases where the agent fails at a later step than the human.
    \item \textbf{AFC}: \textbf{A}gent \textbf{F}ails \textbf{C}orrectly represents the cases where the agent fails at the same step as the human.
\end{itemize}

With this decomposition, we introduce four novel metrics: the Automation Error Rate (AER), the Hallucination Error Rate (HER), the Step Mismatch Error Rate (SMER), and the True Accuracy (TruAcc):

\noindent
\begin{minipage}{0.20\linewidth}
\begin{equation*}
\text{AER} = \frac{\text{AFB}}{\text{TP}}
\end{equation*}
\end{minipage}
\begin{minipage}{0.20\linewidth}
\begin{equation*}
\text{HER} = \frac{\text{AFA}}{\text{TP}}
\end{equation*}
\end{minipage}
\begin{minipage}{0.24\linewidth}
\begin{equation*}
\text{SMER} = \text{AER} + \text{HER}
\end{equation*}
\end{minipage}
\begin{minipage}{0.34\linewidth}
\begin{equation*}
\text{TruAcc} = \frac{\text{AFC} + \text{TN}}{\text{TP} + \text{TN} + \text{FP} + \text{FN}}
\end{equation*}
\end{minipage}\\

The AER rate describes the number of time the agent failed before the human without even having seen the failing step. These errors are likely due to the agent's limitation in exploring the website or validating an assertion.

\emph{Hallucinations} are produced outputs that are coherent and grammatically correct but factually incorrect or nonsensical. In that case, hallucinations would be unlawfully validated step assertions and the HER quantify them.  

The SMER metric helps quantify all the Type III errors by adding the two metrics introduced before. It provides insight into how often the agent arrives at the correct verdict (test case failure) but for the wrong reasons. A low SMER indicates a good alignment between the agent's bug identification process and human evaluation, while a high SMER suggests that the agent may be prone to failures due to its limitations rather than actual bug detection. 

Finally, the TruAcc which, unlike the initially proposed accuracy, takes into account detectable Type III errors of the ATA to represent only verdicts.

%%\subsubsection{Interpretation and Trade-offs}
Based on our defined metrics, an ideal ATA should achieve perfect accuracy, specificity, and sensitivity while maintaining a SMER of zero. However, in practice, trade-offs between these metrics may arise.
We prioritise sensitivity over specificity, as failing test cases should not go unnoticed, whereas false positives can still be manually reviewed. Additionally, SMER serves as a failsafe metric to assess whether high accuracy is genuinely meaningful. A high accuracy score accompanied by a high SMER is problematic, as it indicates that the ATA, despite matching human tester results, does not interpret assertions in the same way as a human tester.

\section{Experiments}
\label{sec:experiments}

\rev{This section presents the evaluation of our benchmark by assessing the two ATAs introduced in Section~\ref{sec:ata}.}
\rev{First, we compare the SeeAct-ATA with PinATA, using GPT-4o, given its strong performance. As expected, our results indicate that PinATA outperforms SeeAct-ATA, achieving a 50\% increase in true accuracy.}
\rev{Next, we focus on PinATA and evaluate its performance across different LLMs (GPT-4o, Sonnet, and Gemini) to analyse their impact. While no major differences are observed, results show a 6\% performance variation between the best and worst-performing models.}
\rev{Finally, we conduct a qualitative analysis of test cases that remain infeasible for PinATA, regardless of the LLM used. This analysis aims to identify key challenges and potential areas for improvement in autonomous test execution.}
\rev{All code and data used in this experiment are available on GitHub~\footnote{Non-Disclosed for double blind review}.}

\subsection{SeeAct-ATA vs PinATA with GPT-4o}

\rev{We evaluated both the SeeAct-ATA and PinATA by executing the 113 test cases from our benchmark across our three target applications.
%All validation experiments were conducted on a machine equipped with an Intel Core i9-9940X CPU @ 3.30GHz (28 logical cores), 64GB DDR4 RAM (4×16GB), and an NVIDIA Quadro P2200 GPU.
}
\rev{To ensure consistency, we reset each system under test to its initial state before every execution, preventing interference from prior runs.
For all executions, both agents were connected to GPT-4o from OpenAI, configured with a temperature of 0 to enhance determinism, though slight variations in responses remained possible.}

\rev{At each execution, the agent produces a verdict, indicating whether the test passes or fails. In the case of a failure, the agent specifies the failed step and identifies the cause—whether it stems from an action or a verification error.}
\rev{Additionally, our execution framework records the entire execution trace of the agent. This trace helps pinpoint the exact step where execution was interrupted if any.}
\rev{The collected data, including verdicts and execution traces, enables the computation of our evaluation metrics, detailed in Table~\ref{tab:resmetric}.}

\begin{table}[ht]
\centering
\small
\setlength{\tabcolsep}{4.5pt}
\begin{tabular}{ll*{7}{S[table-format=1.2]}}
\toprule
& & \multicolumn{7}{c}{\textbf{Evaluation Metrics}} \\
\cmidrule(lr){3-9}
\textbf{Architecture} & \textbf{Application} & {\textbf{Acc}} & {\textbf{Spec}} & {\textbf{Sens}} & {\textbf{AER}} & {\textbf{HER}} & {\textbf{SMER}} & {\textbf{TruAcc}} \\
\midrule
\multirow{4}{*}{\textbf{SeeAct-ATA}} 
& Classified     & 0.50 & 0.33 & 0.67 & 0.30 & 0.15 & 0.45 & 0.20 \\
& Postmill      & 0.56 & 0.67 & 0.44 & 0.13 & 0.07 & 0.20 & 0.47 \\
& OneStopShop   & 0.59 & 0.76 & 0.35 & 0.17 & 0 & 0.17 & 0.53 \\
& Average         & 0.55 & 0.59 & 0.48 & 0.20 & 0.07 & 0.28 & 0.40 \\
\midrule
\multirow{4}{*}{\textbf{PinATA}}
& Classified     & 0.64 & 0.47 & 0.81 & 0.08 & 0.08 & 0.15 & 0.48 \\
& Postmill      & 0.76 & 0.61 & 0.94 & 0.00 & 0.06 & 0.06 & 0.71 \\
& OneStopShop   & 0.73 & 0.62 & 0.90 & 0.09 & 0.02 & 0.11 & 0.63 \\
& Average         & \bfseries 0.71 & 0.57 & 0.88 & 0.06 & 0.05 & 0.11 & 0.61 \\
\bottomrule
\end{tabular}
\captionsetup{
    width=.95\linewidth,      % Width same as table
    justification=centering,
    singlelinecheck=false,  % Ensures centering works for multi-line captions
    font=small,            % Match table font size
    labelfont=bf,          % Bold "Table X"
    textfont=normal        % Normal text for caption
}

\caption{Performance Comparison between SeeAct-ATA and PinATA using GPT-4o.}
\label{tab:resmetric}
\vspace{-10pt}
\end{table}

\rev{Our results demonstrate a clear performance gap between the two ATAs, with PinATA significantly outperforming the SeeAct-ATA. Notably, the true accuracy metric is substantially higher for PinATA, showing an improvement of over 50\% on average (0.61 vs. 0.40).}
\rev{This improvement is largely attributed to the reduction of Type III errors, as reflected in the SMER metric, which is 60\% lower for PinATA (0.11 vs. 0.28). Additionally, PinATA exhibits a much higher sensitivity (0.88 vs. 0.48, an 80\% increase), demonstrating its superior ability to correctly identify failing test cases.}
\rev{Finally, specificity remains comparable between both ATAs, indicating that both agents have a tendancy to generate false failure alerts.}

\subsection{GPT-4o vs Sonnet vs Gemini}

\rev{We leveraged our benchmark to assess the impact of different LLM choices on the design of PinATA. Specifically, we executed the 113 test cases from our benchmark using three commercial high-performing LLMs: GPT-4o, Sonnet, and Gemini (see Table~\ref{tab:eval_arch}). Although additional LLMs could be integrated, our primary goal was to demonstrate the benchmark's usefulness to compare LLMs rather than conducting an exhaustive comparison.}

\begin{table}[ht]
\centering
\begin{tabular}{lccr}
\toprule
\textbf{Backbone LLM} & \textbf{LLM size} & \textbf{Provider} & \textbf{Licence} \\
\midrule
GPT-4o      & 200B*       & OpenAI & Proprietary    \\
  claude-3-5-sonnet  & 175B*       & Anthropic & Proprietary   \\
  gemini-2.0-pro      &  Unknown      & Google & Proprietary    \\ \bottomrule
\end{tabular}
\vspace{5pt}
\captionsetup{
    width=.95\linewidth,      % Width same as table
    justification=centering,
    singlelinecheck=false,  % Ensures centering works for multi-line captions
    font=small,            % Match table font size
    labelfont=bf,          % Bold "Table X"
    textfont=normal        % Normal text for caption
}
\caption{The size of the 3 LLMs used in this experiment ("~*~" are indications, as the exact number of parameters is not confirmed by their providers).}
\label{tab:eval_arch}
\vspace{-20pt}
\end{table}

\rev{Our results reveal only minor differences across the three LLMs (see Table~\ref{tab:llm_comp}), with no single model emerging as significantly superior. True accuracy remains relatively stable across models (0.61 vs. 0.58 vs. 0.62), and the SMER metric exhibits minimal variation (0.11 vs. 0.12 vs. 0.10). The only notable distinction is Sonnet’s superior sensitivity (0.94 vs. 0.88 for GPT-4o and 0.85 for Gemini), indicating a lower rate of undetected failing test cases when using Sonnet. However Sonnet has a lower specificity (0.47 vs 0.57 for GPT-4o and 0.56 for Gemini).}

\begin{table}[ht]
\centering
\small
\setlength{\tabcolsep}{4.5pt}  % Adjust table column spacing
\begin{tabular}{ll*{7}{S[table-format=1.2]}}
\toprule
& & \multicolumn{7}{c}{\textbf{Evaluation Metrics}} \\
\cmidrule(lr){3-9}
\textbf{Model} & \textbf{Application} & {\textbf{Acc}} & {\textbf{Spec}} & {\textbf{Sens}} & {\textbf{AER}} & {\textbf{HER}} & {\textbf{SMER}} & {\textbf{TruAcc}} \\
\midrule
\multirow{4}{*}{\textbf{GPT-4o}} 
& Classified     & 0.64 & 0.47 & 0.81 & 0.08 & 0.08 & 0.15 & 0.48 \\
& Postmill      & 0.76 & 0.61 & 0.94 & 0.00 & 0.06 & 0.06 & 0.71 \\
& OneStopShop   & 0.73 & 0.62 & 0.90 & 0.09 & 0.02 & 0.11 & 0.63 \\
& Average         & \bfseries 0.71 & 0.57 & 0.88 & 0.06 & 0.05 & 0.11 & 0.61 \\
\midrule
\multirow{4}{*}{\textbf{Sonnet}}
& Classified     & 0.60 & 0.27 & 0.93 & 0.07 & 0.07 & 0.14 & 0.47 \\
& Postmill      & \bfseries 0.71 & 0.44 & 1.00 & 0.06 & 0.03 & 0.09 & 0.62 \\
& OneStopShop   & 0.78 & 0.69 & 0.90 & 0.09 & 0.05 & 0.14 & 0.65 \\
& Average         & 0.69 & 0.47 & 0.94 & 0.07 & 0.05 & 0.12 & 0.58 \\
\midrule
\multirow{4}{*}{\textbf{Gemini}}
& Classified     & 0.70 & 0.47 & 0.93 & 0.07 & 0.04 & 0.11 & 0.60 \\
& Postmill      & 0.62 & 0.44 & 0.81 & 0.04 & 0.04 & 0.07 & 0.56 \\
& OneStopShop   & 0.78 & 0.76 & 0.80 & 0.10 & 0.03 & 0.13 & 0.71 \\
& Average         & 0.70 & 0.56 & 0.85 & 0.07 & 0.03 & 0.10 & 0.62 \\
\bottomrule
\end{tabular}
\captionsetup{
    width=.95\linewidth,      % Width same as table
    justification=centering,
    singlelinecheck=false,  % Ensures centering works for multi-line captions
    font=small,            % Match table font size
    labelfont=bf,          % Bold "Table X"
    textfont=normal        % Normal text for caption
}
\caption{Performance Comparison of GPT-4o, Sonnet and Gemini with PinATA.}
\label{tab:llm_comp}
\vspace{-20pt}
\end{table}

% \begin{figure}
%     \centering
%     \includegraphics[width=\textwidth]{figures/TC_results.png}
    
%     \captionsetup{
%     width=.95\linewidth,      % Width same as table
%     justification=centering,
%     singlelinecheck=false,  % Ensures centering works for multi-line captions
%     font=small,            % Match table font size
%     labelfont=bf,          % Bold "Table X"
%     textfont=normal        % Normal text for caption
% }
%     \caption{\textbf{Verdicts of the Models across all Test Cases in the Benchmark.} The table above shows passing tests and the table under shows failing tests. Green/Red mean correct/wrong verdict; Yellow means correct verdict at wrong step.}
%     \label{fig:tc_results}
% \end{figure}

\begin{figure}
    \centering
    \includegraphics[width=\textwidth]{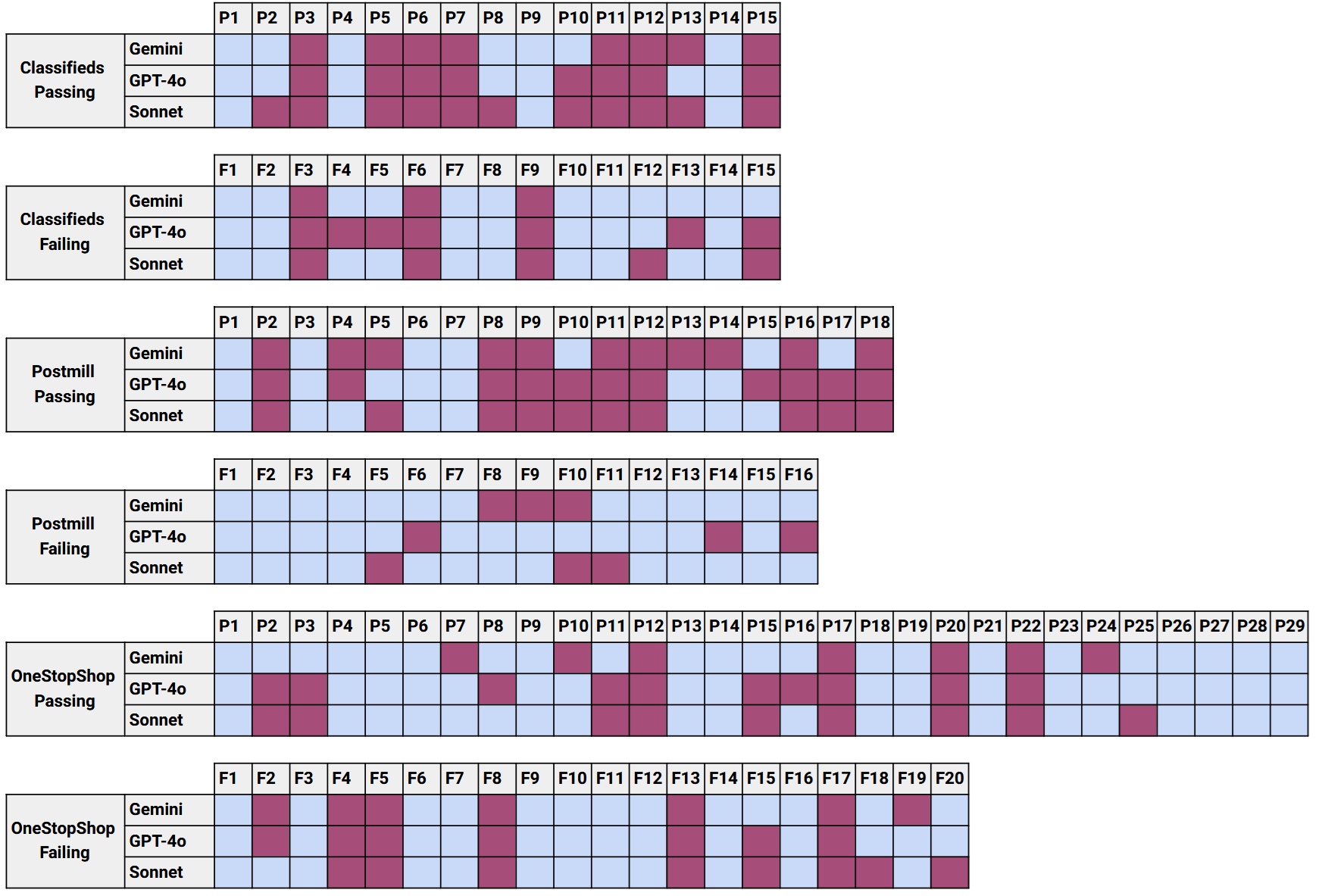}
    
    \captionsetup{
    width=.95\linewidth,      % Width same as table
    justification=centering,
    singlelinecheck=false,  % Ensures centering works for multi-line captions
    font=small,            % Match table font size
    labelfont=bf,          % Bold "Table X"
    textfont=normal        % Normal text for caption
}
    \Description{Executions of the 113 test case with PinATA with three LLMs. Each cell represent an execution by a model. If the cell is in light blue, then the returned verdict is aligned with the one of the human tester. Otherwise, the cell is in purple.}
    \caption{Executions of the 113 test case with PinATA with three LLMs. Each cell represent an execution by a model. If the cell is in light blue, then the returned verdict is aligned with the one of the human tester. Otherwise, the cell is in purple.}
    \label{fig:tc_results_v2}
    \vspace{-10pt}
\end{figure}

Beyond the quantitative results, this experiment highlights that the choice of LLM influences the verdict produced during test case executions. Figure~\ref{fig:tc_results_v2} presents the outcomes of all test case executions across the three selected LLMs, distinguishing between executions that align with the verdict of the human tester (i.e. stopping at the same step with the same verdict) and those that diverge. Notably, our analysis reveals 26 test cases for which PinATA is not aligned with the verdict of the human tester, regardless of the LLM used. These 26 test cases serve as the foundation for our subsequent qualitative analysis, aiming at identifying the underlying challenges and potential areas for improvement.

\subsection{ATA Limitations Taxonomy}
\label{subsec:taxo}

\begin{figure}
    \centering
    \includegraphics[width=0.9\textwidth]{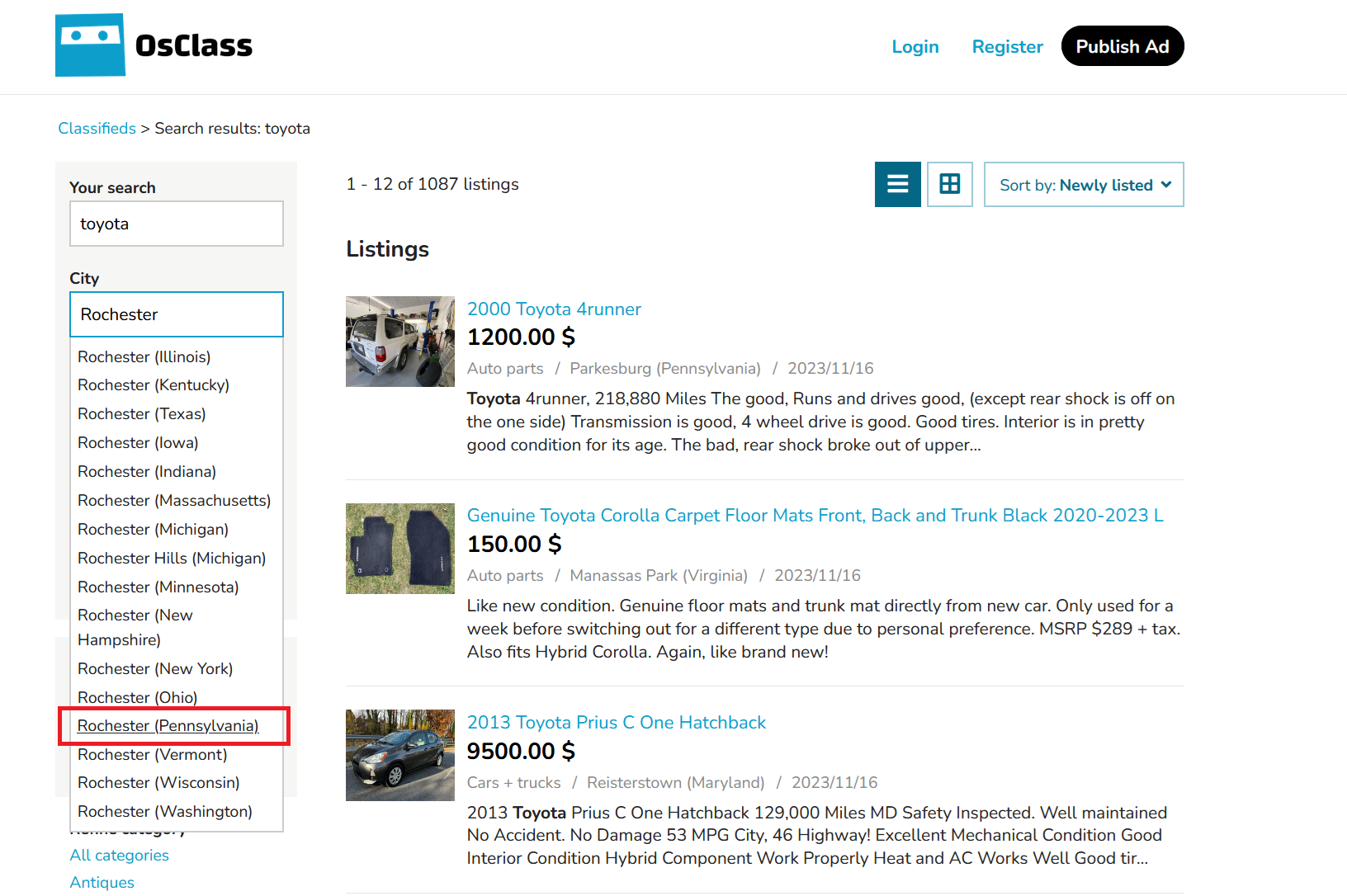}
    
    \captionsetup{
    width=.95\linewidth,      % Width same as table
    justification=centering,
    singlelinecheck=false,  % Ensures centering works for multi-line captions
    font=small,            % Match table font size
    labelfont=bf,          % Bold "Table X"
    textfont=normal        % Normal text for caption
}
    \Description{Screenshot of the Classifieds application showing the completion dropdown menu that need to be used to ensure nominal behaviour.}
    \caption{\textbf{Screenshot of the Classifieds application} showing the completion dropdown menu that need to be used to ensure nominal behaviour.}
    \label{fig:class_screen}
    \vspace{-10pt}
\end{figure}

\rev{To gain deeper insights into the limitations of the ATA, we conducted a qualitative analysis. Our analysis is based on our second experiment, and focused on the executions where, regardless of the selected LLM, the verdict provided by PinATA differed from the one of the professional tester (26 executions out of 113). By isolating these cases, we aimed to identify the intrinsic limitations of PinATA, considering them as potential research directions to advance the field.}

\rev{Two co-authors of this article manually analysed the 26 executions and categorised the errors using an open card-sorting approach~\cite{zimmermann_card-sorting_2016}. In this process, the authors independently grouped errors based on perceived similarities and then reached a consensus on the final set of categories through collaborative discussion. This analysis resulted in five categories: Action Capacity, Action Versatility, User-Interface Observability, Assertion Verifiability and Scenario Conformity. The first three categories illustrate errors also encountered by AWAs, as they stem from challenges in interacting with the web application. In contrast, the last two categories are specific to ATAs.}

\paragraph*{Action Capacity (6 errors out of 26)}

\rev{This category encompasses errors caused by limitations in the technical framework for interacting with the web browser, which restrict the agent's ability to perform certain actions.}
\rev{For instance, the Classified test cases P11 and P12 involve opening a second browser tab, while test case P20 requires accessing the page print options, and test case P7 necessitates opening the browser settings—none of which are currently supported by our agent.}
\rev{Addressing these errors requires a technical framework that provides a broader range of services while ensuring they are easily accessible to the agent.}

\paragraph*{Action Versatility (6 errors out of 26)}

\rev{This category includes errors resulting from the agent's inability to perform complex graphical interface manipulations that cannot be executed directly using the actions provided by the technical framework. For instance, in the Classified application, test case P3 requires entering a location letter by letter to trigger an auto-complete suggestion. However, our agent inputs the entire location at once, disrupting the system and preventing further execution. Figure~\ref{fig:class_screen} illustrates this issue, where the location “Rochester” must be typed incrementally to select the suggestion “Rochester (Pennsylvania).” Another example of the lack of action versatility can be found in F5 of the OneStopShop application where the model does not properly navigate to sub-categories of items due to dropdown menus. Addressing these errors necessitates the development of refined strategies for interacting with the graphical interface. These strategies should dynamically adapt based on the current step and the interface's responses to ensure successful execution.}

\paragraph*{User-interface Observability (5 errors out of 26)}

\rev{This category encompasses errors resulting from the agent's failure to accurately observe the graphical interface, preventing it from recognising the elements it needs to interact with. For instance, in the PostMill test cases P8 and P9, the agent was required to interact with a popup window. However, since popups are not standard HTML elements, our agent—which generates screenshots based on the HTML structure—failed to capture the popup in the screenshot, thereby disrupting the test execution. Another limitation stems from the Set-Of-Mark (SOM) technique we employed, which occasionally lacks precision in distinguishing small UI elements. This issue is evident in the Classified P6 test case, where the technique struggled to accurately identify the necessary elements.}

\paragraph*{Assertion Verifiability (4 errors out of 26)}

\rev{This category encompasses verification errors, which may arise from the agent misunderstanding the verification task, failing to obtain the necessary information, or making an error in analysis. For instance, in test case F9 of the Classified application, the agent was required to verify whether the page displayed elements in a grid format. However, despite the layout not meeting this requirement, the agent incorrectly validated the assertion. Similarly, in test case P11 of application Postmill, the agent is unable to scroll to the bottom of the page to complete its verification. To mitigate these errors, we could enhance the assertor by equipping it with its own specialised action module. It would be capable of not only analysing screenshots but also interacting with elements to confirm their properties. We could also add multi-modal capacities such as DOM analysis. This dedicated module would allow for more thorough and accurate assessments without relying solely on a passive visual analysis.}

\paragraph*{Test Case Conformity (5 errors out of 26)}

\rev{This category includes errors caused by the agent performing unnecessary actions that introduce side effects, resulting in the test not being followed up. As a result, the intended workflow is not tested and actions and/or assertions are not realised. For instance, in F4 of OneStopShop, the wrong assertion of checking for a red heart is not respected by the models and unnecessary navigations to other pages are wrongfully added. Another example can be found in the P17 test case of OneStopShop where the agent anticipates the validation of a popup message, which was supposed to be done in the next step, resulting in a confused orchestrator. Addressing these errors necessitates the development of a more robust planning module that systematically controls the agent’s actions and prevents unintended side effects. For instance, adding future steps in the memory module could limit the anticipation done in some case.}

\subsection{Threats to validity}

In this section, we discuss the threats to validity according to the taxonomy of Wohlin et al.~\cite{wohlin_experimentation_2000}.

\paragraph*{Construct validity}

\rev{We introduced a set of metrics to evaluate to what extent an ATA has similar results compared to a human tester.
However, as we explained in Section~\ref{sec:benchmark}, it is totally possible that an ATA and a human tester obtain a same test verdict by chance (meaning that the ATA has not ensured all the assertions written in the tests), leading to type-3 errors.
To mitigate this phenomenon we introduced four special metrics for the failing tests, based upon a comparison of the exit step.
Unfortunately, this strategy is not possible for the passing tests since the exit step is necessarily the last.
Therefore, the results achieved by ATAs on the passing tests may also subjects to correct verdicts for wrong reasons.}

\paragraph*{Internal validity}

The validity of the test results in our benchmark is the result of manual test execution by the authors.
Even if the authors were very careful performing the manual execution, we cannot guarantee that some tests were not misidentified.

\rev{Our taxonomy of reasons for discrepancies between PinATA and human tester verdicts is derived from qualitative analysis of the ATA's output data, including prompts, responses, and browsing engine logs.
Using this data to determine an underlying reason is a hard task that bear some subjectivity.
We mitigated this subjectivity by proceeding to collaborative tagging sessions, but we cannot guarantee that some reasons were not misidentified.}

\paragraph*{External validity}

\all{The results of the experiments are obtained using a limited number of LLM backbones, fondationnaly trained. The results obtained using fine-tuned LLMs or different language models could be different. In addition, the temperature set to the minimum ensured a reproducibility of our results. However, it might have impacted the quality of the answers of the LLM. A more thorough hyperparameter exploration could improve the generation.}

\rev{While we included web applications from different domains and tests written by professional testers from different companies, the number of tests and applications is still small and therefore our results could be different for other kinds of applications or tests.}

\rev{Our taxonomy of current limitations of the Advanced ATA has been only based on a small set of cases, we cannot guarantee that this taxonomy is complete. Notably, some other limitations could appear using other web application, tests or ATAs}.

\section{Related Work}
\label{rw}

In this section, we present the related work on AWAs and on \rev{scripted test automation.}

\subsection{Scripted Test Automation}

\rev{Among the most prominent techniques in web test automation are scripted automation and keyword-driven testing. Scripted automation involves manually writing test scripts using programming languages or automation frameworks to simulate user interactions with a web application. Similarly, keyword-driven testing structures test cases using predefined keywords representing specific test actions to structure test scripts that can be executed within automation frameworks}~\cite{anupam_automating_2000,garcia_exploring_2024}.
Usually, the testers can add assertions inside the navigation script to ensure the application is correct.
\rev{Automation scripts }are well known to suffer from the test fragility problem~\cite{hammoudi_why_2016,leotta_challenges_2023,memon_chapter_2019}.
Since scripting actions often rely on CSS selectors to find the relevant elements in the web application to interact with, tiny changes in the application design or the browsing context can prevent the script from running and require manual efforts to fix the tests.

Many approaches have tried to mitigate this web test fragility problem.
The first line of approaches aims at automatically repairing a broken test by fixing the broken selector. Historically, the first approaches used the structure of the DOM~\cite{chen_improving_2021,hammoudi_waterfall_2016,brisset_erratum_2022}. Later, some approaches that use visual elements of the page instead of selectors have been introduced~\cite{stocco_visual_2018}. Empirical results show that DOM-based approaches tend to be more robust and easier to evolve~\cite{leotta_visual_2014}.
Another line of approaches targeted the synthesis of robust CSS selectors~\cite{nguyen_generating_2021,leotta_reducing_2014,leotta_using_2015}.
Finally, some approaches focus on the record and replay mechanism to improve its efficiency~\cite{andrica_warr_2011}, for instance by automatically replaying the script during recording to eliminate fragile actions~\cite{long_webrr_2020}.

\subsection{Autonomous Web Agents and Existing ATAs}

The emergence of LLMs and their demonstrated potential as agents in many applications~\cite{wang_survey_2024} has spurred a surge in research efforts aimed at automating web tasks.

\subsubsection{Autonomous Web Agents}

Initial approaches relied on a textual representation of the application, i.e. the Document Object Model (DOM) of web pages, to define the action space for agents. 
However, since the DOM can be considerably large, often exceeding 10,000 tokens, techniques were required to filter out irrelevant information and reduce noise.
Mind2Act consists of a two-stage process where a task specific small language model (SLM) is first involved to rank DOM elements based on their relevance to the task, selecting a subset of candidates~\cite{deng_mind2web_2023}. Then, a LLM such as GPT-4 is queried to predict the next action based on these selected elements.
WebAgent also relies on two specialised LLMs~\cite{gur_real-world_2023}. First, HTML-T5 has been specifically trained to receive the entire DOM and extract the most relevant HTML sections, along with planning for actions. Flan-U-PaLM, a general-purpose LLM, then generates Python programs to perform the actions on the application.
AutoWebGLM forms its observation space by performing optical character recognition (OCR) on a screenshot of the application and combines it with the output of an HTML simplification algorithm~\cite{lai_autowebglm_2024}. The combined observation is fed to a relatively small language model (ChatGLM3-6B) to determine the actions to perform. 
SteP employs a Markov decision process to dynamically compose policies from a library of LLM policies. Each policy provides instructions and examples for a particular subproblem (e.g., searching a list). 

LLMs have recently become multimodal by acquiring vision understanding capabilities. It further reinforced their potential as autonomous agents in web automation, as it augmented their ability to perceive complex web environments.
Although perception is improved, interaction becomes a bit more complex as it requires grounding between the visual and web elements.
With SeeAct, the agent we used as basis for our experiments, the authors experimented with several grounding techniques such as Image annotation (similar to Set-of-marks~\cite{yang_set--mark_2023}), textual choices, and element attributes~\cite{zheng_gpt-4vision_2024}. They report that textual choices is the most efficient grounding method.
WebVoyager combines image annotation and textual choices grounding methods to translate the agent intentions into actions~\cite{he_webvoyager_2024}.
CogAgent takes another route: training a 18B visual language model dedicated to GUI perception and interaction, hence eliminating the grounding problem~\cite{he_webvoyager_2024}. Given a task, the model outputs an action and X,Y coordinates.
VIPACT leverages multi-agent collaboration: an orchestrator agent analyses tasks and coordinates specialised agents and vision expert models to provide detailed visual analysis~\cite{zhang_vipact_2024}.
\rev{Finally, autonomous agents can also be found in mobile applications. For instance, UI-TARS~\cite{qin_ui-tars_2025} is a GUI framework that leverages a native 72B LLM model to execute various tasks and compete with frameworks using OpenAI's GPT-4 model.}

\subsubsection{AI-powered Test Agents}

\all{The literature does not yet contain much information about Autonomous Testing Agents for web applications. This term was first used in the vision paper of Feldt~et~al.~\cite{feldt_towards_2023} where they envision ATAs in a very similar manner as the one defined in our article.}

\rev{GPTDroid~\cite{liu_make_2024} leverages the generation capabilities of LLMs to automate mobile applications GUI testing. It extracts and encodes information into prompts, and then uses the LLM's answers to execute operation within the app, effectively generating activity. These operations are then recorded and saved into a crash report when a bug appears, easing the process of reproducing context. This approach differs from ours as it creates its own actions and expected results from the nominal behaviours of apps the LLM has seen during its training, lacking the precise assertions an expert tester writes in natural language test cases. This type of approach could thus be complementary to ours to discover bugs outside predefined end-to-end test cases.}

\all{Finally, commercial solutions can be found on the internet to automate both test case generation and their executions, such as Flowtest.ai or Octomind.dev. None of these solutions disclose their architecture, nor their effectiveness across applications and test case types.
This lack of transparency motivated us to develop a benchmark that enables end-users to easily compare ATAs architectures, fostering a clearer understanding of their strengths and limitations.
In addition, by providing open-source ATAs we hope to contribute to a more collaborative ATA community.}

% \begin{table}[ht]
% \color{orange}
% \centering
% \begin{tabular}{llcccc}
% \toprule
% \textbf{Architecture} & \textbf{Backbone LLM} & \begin{tabular}[c]{@{}c@{}}\textbf{Test Case}\\ \textbf{Generation}\end{tabular} & \begin{tabular}[c]{@{}c@{}}\textbf{Test}\\ \textbf{Execution}\end{tabular} & \begin{tabular}[c]{@{}c@{}}\textbf{Assertion}\\ \textbf{Verification}\end{tabular} & \textbf{Open-Sourced} \\
% \midrule
% FlowTest.ai & Unknown     &  & \checkmark & \checkmark & \\
% Octomind & Unknown     & \checkmark & \checkmark & \checkmark & \\
% Shortest & Anthropic Claude  &    & \checkmark & &\checkmark    \\
%  \bottomrule
% \end{tabular}
% \vspace{5pt}
%     \captionsetup{
%     width=.95\linewidth,      % Width same as table
%     justification=centering,
%     singlelinecheck=false,  % Ensures centering works for multi-line captions
%     font=small,            % Match table font size
%     labelfont=bf,          % Bold "Table X"
%     textfont=normal        % Normal text for caption
% }
% \caption{\all{Different ATA Providers and Their Characteristics.}}
% \label{tab:rel_ata}
% \end{table}

\section{Conclusion}

\rev{The rapid advancement of Autonomous Web Agents powered by modern LLMs has opened new possibilities for automating web application interactions. Given the resemblance between manual end-to-end test case instructions and the input formats used by AWAs, a compelling question arises: can AWAs be transformed into Autonomous Test Agents capable of executing end-to-end test cases autonomously? This idea has gained traction in industry, with several closed-source ATA solutions emerging. However, little is known about their design or real-world effectiveness.}

\rev{In this article, we explored the distinction between AWAs and ATAs, introduced two ATA implementations—SeeAct-ATA and PinATA—and developed a realistic benchmark for evaluating ATA performance. Our experimental study demonstrated that PinATA can successfully execute 61\% of benchmark test cases with effectiveness comparable to manual testers, even without prior knowledge of the applications. Additionally, when paired with the Sonnet LLM, PinATA achieved an impressive 94\% sensitivity, detecting nearly all failing test cases. However, both SeeAct-ATA and PinATA exhibited a tendency to misclassify passing test cases as failing, with specificity not exceeding 57\%.}

\rev{While these results are promising, several challenges remain. Some limitations are shared with AWAs, meaning ATAs could benefit from advancements in the broader AWA research community. However, ATAs also face unique obstacles, particularly in assertion verifiability and test case conformity. Addressing these challenges will be critical to improving ATA reliability and effectiveness, ultimately paving the way for ATAs to become robust autonomous testers for web application testing.}

\section{Data Availability}

The benchmark, the results of %the modified SeeAct 
\all{the different executions}, the manual verification and the code repository are all openly available on a Zenodo repository~\footnote{\url{https://doi.org/10.5281/zenodo.14937131}}. All the instructions to use the models can be found in a readme at the root of the replication package. 

The dockers for the \all{offline} web applications used for this experiment can be found on the WebArena GitHub repository~\footnote{\url{https://github.com/OSU-NLP-Group/SeeAct}} along with the populated databases. 

\begin{acks}
This work is supported by the company Smartesting.
\end{acks}

\bibliographystyle{ACM-Reference-Format}
\bibliography{bibliography}

\end{document}